\newcommand{\be}{\begin{equation}}
\newcommand{\ee}{\end{equation}}
\newcommand{\ba}{\begin{eqnarray}}
\newcommand{\ea}{\end{eqnarray}}
\newcommand{\bear}{\begin{eqnarray*}}
\newcommand{\eear}{\end{eqnarray*}}

\newcommand{\rf}[1]{(\ref{#1})}

\documentclass[pre,showpacs,twocolumn,superscriptaddress,showkeys]{revtex4}
\usepackage{epsfig,graphicx,psfrag,color}

\begin{document}

\title
{Shared Information   in  Stationary States of Stochastic Processes}
\author{F. C. \surname{Alcaraz}}
\affiliation{ Instituto de F\'{\i}sica de S\~ao Carlos, Universidade de S\~ao
Paulo, \\
Caixa Postal 369, 13560-590, S\~ao Carlos, S\~ao Paulo, Brazil}
\author{V. \surname{Rittenberg}}
\affiliation{Department of Mathematics and Statistics, University of 
Melbourne, Victoria 3010, Australia}
\author{G. \surname{Sierra}}
\affiliation{Instituto de F\'{\i}sica Te\'orica,
UAM-CSIC, Madrid 28049 , Spain}
\date{\today}
\pacs{03.67.Mn, 03.67.-a,02.50.-r,64.60.an,05.40.-a}

\begin{abstract}
We present four estimators of the shared information (or interdepency) 
 in ground-states given that the 
  coefficients appearing in the wavefunction are all real nonnegative numbers 
and therefore can be 
interpreted as probabilities of configurations. Such ground-states 
of hermitian and non-hermitian Hamiltonians can 
be given, for example, 
 by superpositions of valence bond states which can describe 
equilibrium but also stationary states of stochastic models. We 
consider in detail the last case, the system  being a classical not a 
quantum one.
 Using analytical and numerical methods we compare the values of the 
estimators in the directed polymer and the raise and peel models which 
have massive,  conformal invariant and non-conformal invariant massless 
phases. We show that like in the case of the quantum problem, the
estimators verify the area law with logarithmic corrections when  phase transitions take place.
\end{abstract}

\maketitle

 As is well known, in quantum mechanics, for a pure state, if ${\cal C = A + B}$ is 
a bipartition, the von Neumann entanglement entropy $S_q$ is defined as
\be \label{e1}
 S_q ({\cal A}) = S_q ({\cal B}) = -\mbox{Tr}(\rho_{\cal A}\ln\rho_{\cal A})
\ee
where $\rho_{\cal A} = \mbox{Tr}_{\cal B}(\rho)$ and $\rho$ is the density matrix related to the 
ground-state wave function. For one-dimensional spin systems defined by 
Hermitian Hamiltonians, if the lengths $L$ and $l$ of ${\cal C}$ respectively ${\cal A}$, are large one has the area law \cite{ECP}. 
$S_q$ stays finite if the correlation length is finite. If the system is gapless
and conformal invariant, one gets logarithmic corrections 
\be \label{e2a}
  S_q(l,L) \sim \gamma \ln l + C \quad(L>>l), 
\ee
and the finite-size scaling behavior 
\be \label{e2b}
 S_q (l,L) = \gamma \ln (L \sin{(\pi { l}/{L}})/\pi)+ C ,  
\ee
where $\gamma = c/6$ for an open system and $\gamma = 2c/6$ for periodic boundary 
conditions ($c$ is the central charge of the Virasoro algebra) \cite{CAR}. The factor 2 in the latter case 
appears because the systems $A$ and $B$ have two common boundaries.  $C$ is a non-universal constant. The relations \rf{e2a}-\rf{e2b} have been 
checked analytically and numerically for several models \cite{AAA}. 

 In the present paper we consider the shared information  (or interdepency)
 in ground-states 
which are superpositions of valence bond states. 
Our considerations apply to ground-states in which the
coefficients are all real nonnegative and therefore can be interpreted as
probabilities of configurations \cite{bravyi}. 
 This implies that we 
consider the shared information resulting from correlations, in a 
bipartition of a classical and not a quantum system.
The ground-states we study can
describe equilibrium problems 
(the spin $1/2$, $SU(2)$ symmetric one dimensional quantum chains \cite{RAF,ASA} are an example) but also probability distribution functions
(PDF) of stationary states of stochastic processes. 
We are going to concentrate on the latter and therefore also encounter 
 systems which are scale invariant and not conformal 
invariant.
 We will show that if the system is conformal invariant, each 
entanglement estimator $E(l,L)$ behaves like $S_q (l,L)$. The 
constant $\gamma$ has different values for different estimators. If 
the system is scale invariant but not conformal invariant, the  
Eq.~\rf{e2a} stays valid but   the finite-size scaling function 
\rf{e2b}  is different:
\be \label{e2c}
E(l,L) = \gamma \ln(Lg(l/L)), \quad g(x) = g(1-x),
\ee
where  $g(x) \sim \alpha x$ for small $x$ and $C = \gamma \ln\alpha$.  
Like the von Neumann entanglement entropy of quantum systems, the 
estimators detect the existence of long-range correlations.

 We present four estimators of  the shared information and compare them considering
two models defined using the same configuration space. 
We give here only the main results, all the details are going to 
be published elsewhere \cite{TBP}. Exact results are
hard to obtain except for simple cases but for two estimators one can use
Monte Carlo simulations for large system sizes and get reliable results. One of the estimators is not new \cite{CTW,ACL,JLJ}, we are going to 
show its merits and limitations. 

\begin{figure}[t]
\begin{center}
\begin{picture}(150,70)
\put(0,0){\epsfxsize=180pt\epsfbox{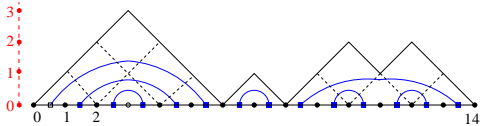}}
\end{picture}
\caption{
(Color online) Example of a link pattern for $L = 14$ and the corresponding Dyck 
path. In the latter there are four contact points and three clusters. 
The shared information is the largest  in the left most cluster.}
\label{fig1}
\end{center}
\end{figure}

 In order to define the configuration space, we consider an open 
one-dimensional system with $L$ sites ($L$ even) connected by $L/2$
 non-intersecting
links (see Fig.~\rf{fig1}). The links can be seen as $U_q(sl(2))$
 (a generalization 
of $SU(2)$ \cite{PAS}) singlets.  There are $C_L = L!/[(L/2)!(L/2+1)!]$ 
configurations of
this kind. There is a bijection between link patterns and restricted 
solid-on-solid (RSOS) configurations also called Dyck paths. A Dyck path is
defined by taking $L+1$ sites situated on the bonds of the link pattern. We
attach to each site $i$ non-negative  integer heights $h_i$ which obey RSOS rules:
\be \label{e3}
 h_{i+1} - h_i =\pm1, \quad h_0 = h_L = 0 \quad  (i = 0,1,\ldots,L).
\ee
The height $h_i$ represents the number of crossed links at the site $i$ 
(see Fig.~\ref{fig1}). If $h_j = 0$, at the site $j$ one has a {\it contact point}.  Between 
two consecutive contact points one has a {\it cluster}. There are four contact
points and three clusters in Fig.~\ref{fig1}. It is easy to see that 
for a 
bipartition, large entanglements take place in large clusters.  
We  present two models. In each of these models one 
has different probabilities for the various Dyck paths. In the cases in 
which one considers stationary states of stochastic models, 
the Dyck paths 
can be seen as an interface between a substrate 
($h_{2i} = 0$, $h_{2i-1} = 1$, 
$i=0,1,\ldots  ,L/2$)) covered by tiles (tilted squares) as shown in 
Fig.~\ref{fig1} and a gas of
tiles (not shown in the figure) \cite{GNPR}. The PDF of the various Dyck paths are determined by the stochastic
process. The latter is defined, in the time-continuous limit, by a
non-hermitian Hamiltonian.
  
 We consider the following two models:

A) {\it The directed polymer model} (DPM) \cite{BEO}. A configuration with 
$m$ contact 
points gets a factor $K^m$ ($K>0$). For $K = 1$, all configurations have the same 
 probability $1/C_L$ and represent the stationary PDF of the Rouse model \cite{ROS} 
of a fluctuating interface. Using reflections about the horizontal axis,
 one can map the 
interface onto a random walker problem.  The walker starts at the 
origin 
and crosses the horizontal axis after $L$ steps. The density of 
 clusters 
is related to the first passage time problem and vanishes like $L^{-3/2}$ for large $L$. In the whole domain 
$0< K< 2$, one is in the same universality class as for $K = 1$. 
For $K = 2$ one gets 
a surface phase transition and for $K > 2$ the density of clusters 
stays 
finite in the thermodynamical limit.

B)  {\it The stationary states of the  raise and peel model} (RPM). 
This is a 
stochastic model \cite{GNPR,AR} in which the adsorption of tiles is local but 
the desorption is non-local. The model has a free parameter $u$. Typical
configurations in the stationary state are shown in  Figs.~ 11 and 15 
of Ref.~\cite{AR}. 
If $0 < u < 1$, the correlation length is finite and one has finite densities of
clusters (see Fig.~12 in \cite{AR}). If $u = 1$, the average density of clusters 
 vanishes in
the thermodynamical limit and the system is conformal invariant (the dynamic
critical exponent $z = 1$). 
 This property makes the model special. 
The valence bonds represent $U_q(sl(2))$ singlets for
$q = \exp(i\pi/3)$. The PDF in the stationary state has also remarkable 
combinatorial properties. For $u > 1$ the system stays critical but conformal
invariance is lost. The exponent $z$ decreases smoothly with $u$ from $1$ to 
zero. 
There are fewer but larger clusters \cite{ALR} than for $u = 1$. 
Because of its rich phase diagram, the RPM is an ideal 
playground to test various estimators. 

A bipartition of the system is obtained in the following way.  
The ensemble of Dyck paths (system $\cal C$ of size $L$ ($L+1$ sites))
 is divided 
into two parts: the sites $0\leq i\leq l$ (part ${\cal A}$) and the sites 
$l\leq j \leq L$ (part 
${\cal B}$). This implies the splitting of each Dyck path which at the site $l$ has 
the height $h_l$ into two ballot paths \cite{KSH}. One RSOS path which 
starts at $i = 0$ and ends at the site $l$ at the height $h_l$ and another one 
which starts at $i=l$, with height $h_l$,  and ends at a height zero at $i = L$. We denote by  
 $ P(a(h_l),b(h_l))$ the
probability to have a given Dyck path in ${\cal C}$ formed by the ballot paths
  $a(h_l)$($b(h_l)$)  in ${\cal A}$, respectively in $B$. We 
consider the marginals   
\be \label{e5}
 P(a(h_l)) = \sum_b P(a(h_l),b(h_l)),
\ee
and $P(b(h_l))$.
 The probability to have a height $h_l$ at the site $l$ is
\be \label{e6}
 F(h_l) = \sum_a P(a(h_l)) = \sum_b P(b(h_l)).
\ee 

 We present the four estimators. They all measure in different ways the 
amount of information that can be obtained about the ballot paths in $B$ if one
observes the ballot paths in ${\cal A}$.

I) {\it Mutual Information.}
\be \label{e7}
 I(l,L) =- \sum_{h_l, a(h_l),b(h_l)} P(a(h_l),b(h_l)) \ln{\frac{P(a(h_l),b(h_l))}{P(a(h_l))P(b(h_l))}}.
\ee
This is a known estimator \cite{CTH}. 

II) {\it Boundary Shannon Entropy }
\be \label{e8}
 S(l,L) = H(L) - H(l) - H(L-l) ,    
\ee
where $H(M) = - \sum_k P_k\ln(P_k)$ is the Shannon entropy for a system of 
size $M$ and  $P_k$ is the probability to have a Dick path $k$. Notice that if, like in model $A$ with $K=1$, all configurations have the same 
probabilities and their number is $Z(M)$,
\ba \label{e9}
&& S (l,L) = -\ln Q(L,l), \nonumber \\
&& Q(L,l) = Z(l)Z(L-l)/Z(L) = F(h_l=0),
\ea 
where $Q(L,l)$ is the probability to have the two systems ${\cal A}$ and ${\cal B}$ separated 
by a contact point at the site $l$.

III) {\it Density of Contact Points Estimator} 

\be \label{e10}
D (l,L) = \ln(1/F(h_l=0)).
\ee
Notice that $D(l,L)$ and $S(l,L)$ coincide if all configurations have the same 
probabilities.  The physical meaning of $D(l,L)$ is simple: if ${\cal A}$
 and ${\cal B}$ have 
a small probability to be separated by a contact point  
the estimator is 
large. This should be the case since  the shared information among 
${\cal A}$ and ${\cal B}$ is large. 

 In the continuum, $F(h_l = 0)$ can be replaced by $\rho(l,L)$, the 
local density of contact points at the  distance $l$ from the origin for a
system of size $L$. This is an average of a local operator. Let us observe 
that for $1<<l,L$  the density $\rho$ stays finite, and therefore  
 $D(l,L)$ is also finite. 
If for large 
values of $l$ and $L$ one has,   
\be \label{e11}
 \rho(l,L) = 1/[Lg(l/L)]^X,
\ee
with  $g(x)\sim  \alpha x$ 
for small $x$ ($\alpha$ being a constant), 
one obtains \rf{e2a}  and \rf{e2c}
 with $\gamma=X$. 
 [The average number of clusters is $\rho(L) \sim L^{-X}$.] 

IV) {\it Valence Bond Entanglement Entropy.}

 This estimator was introduced independently by Chhajlany et al 
\cite{CTW} and Alet et al \cite{ACL} and further studied by Jacobsen and Saleur 
\cite{JLJ} (see also \cite{MMA}). The estimator is the average height at the site $l$, 
for a system of size $L$:  
\be \label{e13}
 h(l,L) = \sum_{h_l} h_l F(h_l).
\ee

 We give the main  results for the four estimators for each of the two models 
presented above (see \cite{TBP}). 
 
{\bf Directed Polymer Model:}

It is easy to show that for $K=1$ one has
\ba \label{e13p} 
 I(l,L) &=& 1/2\ln(l(L-l)/L) + \gamma_e -1/2(\ln(\pi/2)+1) =
\nonumber \\
&\sim& 
1/2 \ln(l) + 0.303007 \quad  (L >>l),\\
\label{e13pp}
 S(l,L) &=& D(l,L)=\frac{3}{2}\ln(\frac{l(L-l)}{L}) + \frac{1}{2}\ln {
\frac{\pi}{8}}.
\ea
In \rf{e13p} $\gamma_e$ is the Euler constant. 
Notice that Eq.\rf{e2a}
stays valid, the 
finite-size  scaling functions in \rf{e13p} and \rf{e13pp} are the same but 
different from the one given by \rf{e2b}.
We have checked \cite{TBP} that 
except for the additive constants, Eqs. \rf{e13p} and \rf{e13pp} are valid in the 
whole interval $0 < K < 2$ as expected from universality. The 
valence bond entanglement entropy does not get logarithmic 
but power corrections to the area law since $h(l,L) \sim  l^{1/2}f(l/L)$ 
in the whole interval $0 < K < 2$ 
 \cite{BEO}. For $K > 2$ and large values of $L$, the 
density of clusters is finite and all estimators verify the area law. 

{\bf Raise and Peel Model:}

\noindent ${\bf u < 1}:$

$D(l,L)$ and $h(l,L)$ are finite since the average density of clusters is
finite. $I(l,L)$ and $S(l,L)$ were not computed.

\noindent {\bf u=1 } ({\it conformal invariance}):

$I(l,L)$ and $S(l,L)$ are hard to obtain since the PDF is know exactly only 
for small lattices. Some rough estimates  given in \cite{TBP} 
show that they are compatible with 
\rf{e2b}.

$D(l,L)$ is obtained in the following way. As shown in \cite{APR} in an "almost" 
 rigorous way, the density of contact points is the average of a local operator of 
 a conformal field theory and has the expression: 
$\rho (l,L) = m/[L\sin(\pi l/L)]^{1/3}$ where  $ m=
-\sqrt{3} \Gamma(-{1}/{6})/(6\pi^{5/6})$.
Using \rf{e11} one finds:
$D(l,L) = 1/3 \ln(L\sin(\pi l/L)/\pi) + 0.28349.$ 
This is precisely Eq.~\rf{e2b} in which $\gamma = 1/3$ is not given by  the 
central charge of the Virasoro algebra (one would expect $\gamma = 1/6$ if this 
would have been the case) but by the scaling dimensions of a local 
operator. Moreover one can estimate what would happen if the segment ${\cal A}$  
would be inside an infinite system (two separation points). 
The estimator $D(l,\infty)$ is given by the two-point correlation function of 
the densities separated by $l$, measured in  \cite{ALR}. One obtains a violation of the area law \rf{e2a} with $\gamma = 2/3$.

$h(l,L)$ was obtained using Monte Carlo simulations for lattices up to $L = 9600$.
The results are compatible with Eq.~\rf{e2b}:
$h(l,L) = 0.277 \ln(L \sin(\pi l/L)/\pi) + 0.73$.
\begin{figure}[t]
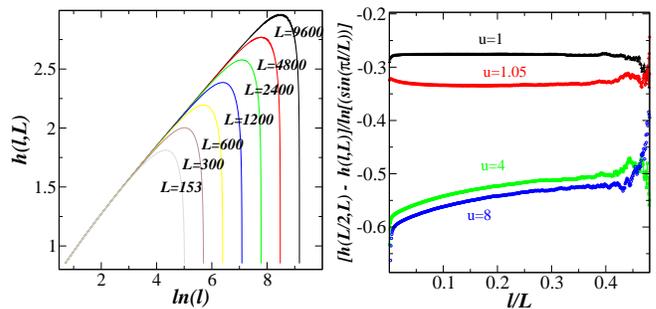

\centering{ \includegraphics [angle=0,scale=0.29]
{fig2a.eps}\hspace{-3.0cm}\hfill\includegraphics
[angle=0,scale=0.29] {fig2b.eps} } \caption{
(Color online). Left plot: The estimator $h(l,L)$ for $u=1$ for 
various lattice sizes L
($l<<L$) as a 
function of $\ln l$. 
 Right plot: The scaling function $[h(L/2,L) -h(l,L)]$ divided by $\ln(\sin(\pi l/L))$ 
for various values of $u$ measured on a lattice of size $L = 2400$. For $u = 1$ 
one should get a constant independent on $l$. }
\label{fig2}
\end{figure}
 These results were obtained in the following way. First  we have taken $l<<L$
and plotted $h(l,L)$ as a function of $\ln l$ for various values of $L$ (see 
Fig.~\ref{fig2}). One can see that there is a domain where we have a straight line 
which is $L$ independent. This has allowed to get $\gamma$ and $C$ (see Eq.~
\rf{e2a}). 

 Next, we have considered the quantity $[h(L/2,L) - h(l,L)]/\ln[\sin(\pi l/L)]$ 
for various values of $L$. If Eq.~\rf{e2c} is valid one should obtain a constant 
equal to $- \gamma$. The data are presented in Fig.~\ref{fig2} for $L = 2400$ and one can see 
that this is indeed the case. We should mention that considering periodic 
boundary conditions and the boundary Coulomb gas formalism, Jacobsen and 
Saleur \cite{JLJ} have obtained the value of $\gamma$ (one has to take half of their value
since we deal with an open system) $\gamma = \sqrt{3}/2\pi \sim 0.275$ which is 
compatible with our result.
 It is remarkable that the two estimators $D(l,L)$ and $h(l,L)$ give 
 values for 
$\gamma$ which are close to each other. 

\noindent ${\bf u > 1}:$

 The estimator $D(l,L)$ can be computed using Monte Carlo simulations. A 
rough estimate of $\gamma$ can be obtained using the equality $\gamma = X$ where the 
exponent $X$ is related to the density of clusters (see \cite{PAS}).
The exponent $X$ varies between $1/3$ and $1$ when $u$ increases from 
$1$ to large values. 
One has $X=0.5$, $0.6$ and $0.85$ for $u=1.2$, $1.5$ and $10$, 
respectively \cite{ALR}.

We have done a more detailed study for $u = 4$ ($z \sim 0.3$) in this 
case and 
for $D(l,L)$ we found 
 \rf{e13p} with $\gamma = 0.73 \pm 0.03$ and a scaling function 
$g(l/L)$ different of \rf{e2b} (conformal invariance is lost at $u = 4$). 
We have 
also studied $h(l,L)$ and found $\gamma = 0.63 \pm 0.03$ and a function 
$g(x,L)$ (see right plot of Fig.~\ref{fig2}) 
 equal
within errors to the one observed for $D(l,L)$ \cite{FN1}.
 Notice that for both 
estimators the values of $\gamma$ have increased by more than a factor of two 
as compared with the values observed at $u = 1$. An increase 
of the shared information was expected 
since there are larger  clusters connecting the subsystems ${\cal A}$ 
and ${\cal B}$.

The estimators defined above can be used for stationary states of other 
processes not taking place in the Dyck paths configuration space. A simple 
example is the asymmetric exclusion problem (ASEP) with a density 
$r$ of particles on a ring of perimeter $L$. 
The role of the heights in the Dyck paths is played by the deviation of 
the number of particles in a subsystem (size $l$) from the number $rl$. The 
system being critical, one expects corrections to the area law. One finds 
indeed:
\ba \label{ASEP}
&& I(l,L) = S(l,L)=D(l,L) = \nonumber \\
&&1/2\ln(l(L-l)/L)+ 1/2\ln(2\pi r(1-r)).
 \nonumber 
\ea
Notice that $\gamma$ in \rf{e2a} is $r$ independent.

 We have shown that the four estimators of the shared information between 
two subsystems, defined above, verify the area law. If the system is 
gapless, one obtains (with one exception) logarithmic corrections with 
the 
coefficient $\gamma$ in \rf{e2a} increasing if the shared information is larger. 
The exception is the average height, which in the directed polymer 
model 
gets power corrections probably due to the fact that the density of 
clusters decreases very fast with the size of the system. As a result, the 
existence of corrections to the area law can be used to detect the existence
of phase transitions. 
Moreover, the observation of a finite-size scaling 
law like \rf{e2b} can be an indication of conformal invariance. 
 The estimators presented here have been generalized to the 
multi-partition case (see [7]).

\section*{Acknowledgments}
We would like to thank P. Pyatov for related discussions.
 The work of F. C. A. was
partially supported by FAPESP and CNPq (Brazilian Agencies) and the one 
of V. R. was supported  by ARC and DFG. 
F.C.A. and V.R. thanks the warm hospitality 
of the Instituto de F\'{\i}sica Te\'orica, 
UAM-CSIC, Madrid, Spain, where part of this work was done.


\begin{thebibliography}{99}
\bibitem{FN1} 
 For $u > 1$, the average heights of the raise and peel model increase logarithmically with $L$. This point 
was missed in \cite{AR}, where the slow logarithmic effects were 
overlooked  and it 
was claimed that the the heights reach a constant value for large values of $L$.
\end{thebibliography}
\end{document}